\documentclass[12pt]{article}
\usepackage{latexsym}
\usepackage{amssymb,amsfonts,amsmath}
\usepackage{graphicx} 
\usepackage{indentfirst}
\usepackage{bbm}
\usepackage{amssymb}
\usepackage{verbatim}
\usepackage{amsmath, amsthm,amssymb}
\usepackage{mathrsfs}
\usepackage{hyperref}
\usepackage{amsfonts}
\usepackage{dsfont}

\newcommand {\cD}{{\cal D}}
\newcommand {\cE}{{\cal E}}

\newcommand {\cN}{{\cal N}}

\newcommand {\cR}{{\cal R}}

\newcommand {\cT}{{\cal T}}

\newcommand {\cV}{{\cal V}}

\newcommand {\cX}{{\cal X}}
\newcommand {\cY}{{\cal Y}}

\def\a{\alpha}
\def\b{\beta}

\def\d{\delta}

\def\g{\gamma}
\def\G{\Gamma}

\def\j{\psi}
\def\k{\kappa}
\def\l{\lambda}

\def\o{\omega}

\def\q{\theta}
\def\r{\rho}
\def\s{\sigma}

\def\z{\zeta}

\def\F{\Phi}
\def\J{\Psi}
\def\L{\Lambda}

\def\S{\Sigma}
\def\U{\Upsilon}

\def\ri{{\rm i}}

\newcommand{\ad}{{\dot{\alpha}}}                           
\newcommand{\bd}{{\dot{\beta}}}                            

\newcommand{\pa}{\partial}                           
\newcommand{\hf}{\frac12}

\newcommand{\be}{\begin{equation}}
\newcommand{\ee}{\end{equation}}
\newcommand{\bea}{\begin{eqnarray}}
\newcommand{\eea}{\end{eqnarray}}
\newcommand{\non}{\nonumber}
\newcommand{\ba}{\begin{array}}
\newcommand{\ea}{\end{array}}

\def\double #1{#1{\hbox{\kern-2pt $#1$}}}

\newcommand{\sU}{\mathsf{U}}

\newcommand{\bsubeq}{\begin{subequations}}
\newcommand{\esubeq}{\end{subequations}}

\newcommand{\rd}{\mathrm d}
\numberwithin{equation}{section}

\begin{document}

\begin{center}
{\Large \bf 
Taking a vector supermultiplet apart: Alternative Fayet-Iliopoulos-type terms}
\end{center}

\begin{center}
{\bf Sergei M. Kuzenko} \\
\vspace{5mm}

\footnotesize{
{\it Department of Physics M013, The University of Western Australia\\
35 Stirling Highway, Crawley W.A. 6009, Australia}}  
~\\
\vspace{2mm}
\end{center}

\begin{abstract}
\baselineskip=14pt
Starting from an Abelian $\cN=1$ vector supermultiplet $V$ coupled to 
conformal supergravity, we construct from it a nilpotent real scalar Goldstino superfield
$\mathfrak V$ of the type proposed in arXiv:1702.02423. 
It contains only two 
independent component fields, the Goldstino and the auxiliary $D$-field.  
The important properties of this Goldstino superfield are: (i) 
it is gauge invariant;  and (ii) it is super-Weyl invariant. 
As a result, the gauge prepotential 
can be represented as $V= {\cal V} +\mathfrak V$, where $\cal V$ 
contains only one independent component field, 
modulo gauge  degrees of freedom,  which is the gauge one-form.
Making use of $\mathfrak V$ allows us to introduce  new
Fayet-Iliopoulos-type terms, which differ from the one proposed in
arXiv:1712.08601 and share with the latter the property that  
gauged $R$-symmetry is not required.
\end{abstract}

\renewcommand{\thefootnote}{\arabic{footnote}}


\section{Introduction}

In quantum  field theory with a symmetry group $G$ spontaneously broken
to its subgroup $H$, the multiplet of matter fields transforming according 
to a linear representation 
of $G$ can be split into two subsets: (i) the massless Goldstone fields; and 
(ii) the other  fields that are massive in general. 
Each subset transforms nonlinearly with respect to $G$
and linearly under $H$. Each subset may be realised in terms of constrained 
fields transforming linearly under $G$ \cite{Kibble,CWZ}.
In the case of spontaneously broken supersymmetry \cite{VA}, 
every superfield $U$ containing the Goldstino may be split 
into two supermultiplets, 
one of which is an irreducible Goldstino superfield\footnote{The notion of irreducible and reducible Goldstino
 superfields was introduced in \cite{BHKMS}. For every irreducible Goldstino superfield,
 the Goldstino is its only independent component. 
 Reducible Goldstino superfields also contain auxiliary field(s) 
 in addition to the Goldstino.
 }   
and the other contains the remaining component fields \cite{BHKMS},
in accordance with the general relation between linear and nonlinear realisations of 
$\cN=1$ supersymmetry \cite{IK}. 
It is worth recalling the example worked out in \cite{BHKMS}. 
Consider the irreducible chiral Goldstino superfield  
 $\cX$, $\bar D_\ad \cX=0$, introduced in \cite{IK,Rocek}.
It  is defined to obey the
 constraints \cite{Rocek} 
 \bea
 \cX^2 =0~, \qquad  {f}\cX = -\frac 14  \cX \bar D^2 \bar \cX,~
 \eea
 where $f$ is a real parameter 
  characterising the scale of supersymmetry breaking. 
  As $U$ we choose the reducible chiral  Goldstino superfield 
  $X$,  $\bar D_\ad X=0$, proposed in \cite{Casalbuoni,KS}. 
  It is subject only  to the constraint
 \bea
 X^2=0~.
 \eea
It was shown in \cite{BHKMS} that $X$ can be represented in the form
\bea
X = \cX +\cY~,\qquad 
{f} \cX := - \frac{1}{4} \bar D^2 (\bar \S \S)\, , \qquad 
\S :=  - 4 f \frac{\bar X}{\bar D^2 \bar X}~,
\label{1.3}
\eea
where the auxiliary field $F$ of $X$ is the only independent component of the 
chiral scalar $\cY$. Originally, 
the irreducible Goldstino  superfield  $\S$
 was introduced in \cite{KTyler} to be a
modified complex linear superfield, $-\frac 14 {\bar D}^2\Sigma={f}$, 
which is nilpotent and obeys a holomorphic nonlinear constraint,
\bea
\S^2=0~, \qquad
{f}D_\alpha\Sigma = -\frac 14 \Sigma{\bar D}^2D_\alpha\Sigma~.
\eea
These properties follow from \eqref{1.3}.

The approach advocated in \cite{BHKMS} may be pursued one step further
with the goal to split any unconstrained superfield $U$ into two supermultiplets,
one of which is a reducible Goldstino supermultiplet. This has been implemented 
in \cite{CDF} for the reducible chiral Goldstino superfield $X$.
There exist two other reducible Goldstino superfields: (i) the three-form 
variant of $X$ \cite{FKRR,BK17}; and (ii) the nilpotent real scalar superfield
introduced in \cite{KMcAT-M}. In the present paper we make use of (ii)
in order to split a $\sU(1) $ vector supermultiplet into two constrained superfields.
Our construction makes it possible to introduce  new
Fayet-Iliopoulos-type terms, which differ from the one recently proposed in
 \cite{CFTV}  and share with the latter the property that 
 gauged $R$-symmetry is not required.

In this paper, we make use of the simplest formulation for $\cN=1$ conformal 
supergravity in terms of the superspace geometry of \cite{GWZ}, 
which underlies the Wess-Zumino approach \cite{WZ}
to old minimal supergravity \cite{old1,old2}.  
This approach requires the super-Weyl transformations of \cite{HT} 
(defined in the appendix)
to belong to the supergravity gauge group. Our notation and conventions follow 
\cite{Ideas}.


\section{Constructing a Goldstino superfield}

Consider a massless vector supermultiplet in a conformal supergravity background.
It is described by a real scalar prepotential $V$  defined modulo gauge transformations
\bea
\d_\l V = \l +\bar \l~, \qquad \bar \cD_\ad \l =0~.
\label{2.1}
\eea
As usual, the prepotential is chosen to be super-Weyl inert, $\d_\s V=0$. 
In what follows, 
we assume that the top component ($D$-field) of $V$ 
is nowhere vanishing. 
In terms of  the gauge-invariant field strength \cite{WZ}
\bea
W_\a := -\frac{1}{4} (\bar \cD^2 - 4R) \cD_\a V~,\qquad \bar \cD_\bd W_\a=0~,
\label{W}
\eea
our assumption means that
the real scalar
$\cD W:=\cD^\a W_\a =\bar \cD_\ad \bar W^\ad$ is nowhere vanishing.

It is instructive to consider a simple supersymmetric gauge theory in which 
the above assumption is compatible with the equations of motion.  
Within the new minimal formulation for $\cN=1$ supergravity \cite{new,SohniusW3}, 
the dynamics of the massless vector supermultiplet 
with a Fayet-Iliopoulos (FI) term \cite{FI}
is governed by the gauge invariant and super-Weyl invariant action
(see, e.g., \cite{FGKV})
\bea
S[V] =   \int \rd^4 x \rd^2 \q  \rd^2 \bar{\q} \, E\,\Big\{
\frac{1}{16} V \cD^\a (\bar \cD^2 -4R ) \cD_\a V- 2f { L} V\Big\}~,
\label{2.2}
\eea
where  $ L$ is  the conformal compensator for new minimal
supergravity \cite{deWR} (and as such $L$ is nowhere vanishing).
It  is a  real covariantly linear scalar superfield,
\bea  
(\bar \cD^2 -4R) { L}  =0~, \qquad \bar { L}= { L}~,
\label{2.3}
\eea
with the super-Weyl transformation $\d_\s { L} = (\s+\bar \s) { L}$.
The second term in the action is the  FI term, with $f$ a real parameter. 
The equation of motion for $V$ is  
$\cD W=-4f  L$, and it implies
that $\cD W$ is indeed nowhere vanishing.

Since $\cD W$ is nowhere vanishing, we can introduce 
(as an extension of the construction in section 5.2 of \cite{KMcAT-M}) 
the following  scalar superfield  
\bea
{\mathfrak V} := - 4 \frac{W^2 \bar W^2}{(\cD W)^3}~, \qquad 
W^2 := W^\a W_\a~. 
\label{2.5}
\eea
This superfield is gauge invariant, $\d_\l {\mathfrak V} =0$,  and super-Weyl
invariant, 
\bea
\d_\s {\mathfrak V} =0~,
\eea
as follows from the super-Weyl transformation laws of $W_\a$ and $\cD W$:
\bea
\d_\s W_\a = \frac{3}{2} \s W_\a~, \qquad \d_\s \cD W = (\s + \bar \s) \cD W~.
\eea
By construction, it obeys the following nilpotency conditions
\begin{subequations} \label{2.6}
\bea
{\mathfrak V}^2&=&0~, \label{2.6a}\\
{\mathfrak V} \cD_A \cD_B {\mathfrak V} &=&0~,\label{2.6b} \\
{\mathfrak V} \cD_A \cD_B \cD_C {\mathfrak V} &=&0~, \label{2.6c}
\eea
\end{subequations}
which mean that $\mathfrak V$ is the Goldstino superfield introduced 
in \cite{KMcAT-M}.\footnote{The Goldstino superfield constrained by \eqref{2.6}
contains only two independent fields, the Goldstino and the auxiliary $D$-field.   
This can be shown by analogy with the $\cN=2$ analysis in 
three dimensions \cite{BHKT-M}.}
Associated with $\mathfrak V$ is the 
the covariantly chiral spinor ${\mathfrak W}_\a $ which is obtained from \eqref{W} 
by replacing $V$ with $\mathfrak V$.
As shown in \cite{KMcAT-M}, the constraints \eqref{2.6} imply that 
\bea
{\mathfrak V} := - 4 \frac{ {\mathfrak W}^2 \bar {\mathfrak W}^2}{(\cD {\mathfrak W})^3}~.
\label{2.99}
\eea
Choosing $V=\mathfrak V$ in \eqref{2.2} gives the Goldstino superfield action 
proposed in \cite{KMcAT-M}.

In order for our interpretation of $\mathfrak V$ as a Goldstino superfield
to be consistent, its 
$D$-field should be nowhere vanishing, which is equivalent to the requirement that  
$\cD {\mathfrak W}$ be nowhere vanishing. As follows from \eqref{2.5}, this condition 
implies that $\cD^2 {W}^2$ is nowhere vanishing.  
 To understand what the latter implies, let us  introduce the component fields of the vector supermultiplet following \cite{KMcC}
\bea
W_{\a}\arrowvert = \j_{\a}~,\qquad
-\frac{1}{2}\cD^{\a}W_{\a}\arrowvert=D~,
\qquad
\cD_{(\a}W_{\b)}\arrowvert 
=2 {\rm i} {\hat F}_{\a\b}
&=& {\rm i} (\s^{ab})_{\a\b}{\hat F}_{ab}~,
\eea
where  the bar-projection, $U|$, means switching off the superspace Grassmann variables, and 
\bea
\label{eq:F_ab defn}
{\hat F}_{ab} &=& F_{ab} -
\frac{1}{2}(\J_{a}\s_{b}{\bar \j} + 
\j\s_{b}{\bar \J}_{a}) +
\frac{1}{2}(\J_{b}\s_{a}{\bar \j} + 
\j\s_{a}{\bar \J}_{b})~,
\non\\
F_{ab} &=& \nabla\!_{a}V_{b} 
- \nabla\!_{b}V_{a} 
- {\cT_{ab}}^{c}V_{c}~,
\eea
with $V_a= e_a{}^m (x) \,V_m (x)$  
the  gauge  one-form, and $\J_a{}^\b$ the gravitino. 
The operator $\nabla_a$ denotes a spacetime covariant derivative with torsion,
\be
\label{eq:covariant derivative algebra}
\left[\nabla\!_{a}, \nabla\!_{b}\right] = {\cT_{ab}}^{c}\,\nabla\!_{c}
+ \frac{1}{2}\,\cR_{abcd} M^{cd}~,
\ee
where $\cR_{abcd}$ is the curvature tensor and $\cT_{abc}$ 
is the torsion tensor. The latter is related to the gravitino by
\be
\cT_{abc} = -\frac{\rm i}{2}(\J_{a}\s_{c}{\bar \J}_{b}
- \J_{b}\s_{c}{\bar \J}_{a})~.
\ee
For more details, see \cite{Ideas,KMcC}.
We deduce from the above relations that 
\bea
-\frac{1}{4} \cD^2 {W}^2| = D^2 - 2 F^{\a\b} F_{\a\b} +
\text{fermionic terms}~.
\eea
We conclude that the electromagnetic field should be weak enough 
to satisfy
\bea
D^2 - 2 F^{\a\b} F_{\a\b}  \neq 0~,
\label{2.15}
\eea
in addition to the condition $D \neq 0$.
The $D$-field of $\mathfrak V$  is
\bea
-\hf \cD {\mathfrak W} | = D \Big| 1 - 2 \frac{F^{\a\b} F_{\a\b}}{D^2} \Big|^2
+\text{fermionic terms}~.
\eea

Making use of the Goldstino superfield ${\mathfrak V}$ leads
to a  new parametrisation for the gauge 
prepotential given by 
\bea
V = \cV +{\mathfrak V}~.
\eea
It is $\cV$ which varies under the gauge transformation \eqref{2.1},
$\d_\l \cV = \l + \bar \l$, while $\mathfrak V$ is gauge invariant by construction.
Modulo purely gauge  degrees of freedom, $\cV$ contains only one independent 
field, which is the gauge one-form.

There exists a different way to construct 
a reducible Goldstino superfield in terms of $V$, which is  given by 
\bea
\hat {\mathfrak V} := - 4 \frac{W^2 \bar W^2}{\cD^2 W^2 \bar \cD^2 \bar W^2}
\cD W~.
\label{2.17}
\eea
Unlike $\mathfrak V$ defined by \eqref{2.5}, 
this gauge-invariant superfield is not manifestly super-Weyl invariant. 
Nevertheless, it
proves to be invariant under the super-Weyl transformations, 
\bea
\d_\s \hat {\mathfrak V} =0~, 
\eea
as follows from the observation \cite{CF}
(see also \cite{KMcC}) 
that 
\bea
\Big(\cD^2 -4\bar R\Big) \frac{W^2}{\U^2}
\eea
is super-Weyl invariant for any compensating 
(nowhere vanishing) 
real scalar $\U$ with the super-Weyl transformation law
\bea
\d_\s \U = (\s+ \bar \s) \U~.
\eea
In the new minimal supergravity, we can identify 
\begin{subequations}\label{2.22}
\bea
 \U =L~. \label{2.22a}
 \eea
In the case of the old minimal formulation for $\cN=1$ supergravity
\cite{WZ,old1,old2}, we choose
\bea
\U = \bar \F \F~, \qquad \bar \cD_\ad \F =0~,
\label{2.22b}
\eea
\end{subequations}
where   the chiral compensator $\F$
has the super-Weyl transformation law 
$\d_\s \F = \s \F$.

By construction, the superfield \eqref{2.17} obeys the nilpotency conditions \eqref{2.6}.
It may be shown that the composites \eqref{2.5} and \eqref{2.17} coincide 
if $V$ is chosen to be $ {\mathfrak V}$ 
or $\hat {\mathfrak V}$.
Thus the two Goldstino   superfields $ {\mathfrak V}$ or $\hat {\mathfrak V}$ 
differ only in the presence of a gauge field.
It follows from the definition \eqref{2.17} that $\hat {\mathfrak V}$ is well defined 
provided the condition \eqref{2.15} holds. 
The same definition tells us that the $D$-field of $\hat {\mathfrak V}$ is equal to 
\bea
-\hf \cD \hat {\mathfrak W} | = D +\text{fermionic terms}~.
\eea

The composite \eqref{2.17} was introduced 
in a recent paper \cite{CFTV}. 
The authors of \cite{CFTV} put forward the supersymmetric invariant 
\bea
\hat{\mathfrak I}_{\rm FI} =   \int \rd^4 x \rd^2 \q  \rd^2 \bar{\q} \, E\,
\U \hat {\mathfrak V}
\label{2.24}
\eea
as a novel FI term that does not require gauged $R$-symmetry.
The compensating superfield $\U$ was chosen in \cite{CFTV} to be 
the old minimal expression \eqref{2.22b}.

We propose an alternative FI-type invariant
\bea
{\mathfrak I}_{\rm FI} =   \int \rd^4 x \rd^2 \q  \rd^2 \bar{\q} \, E\,
\U  {\mathfrak V}~.
\label{2.25}
\eea
It also does not require gauged $R$-symmetry.
In addition, it does not require  \eqref{2.15}.

Actually, the above constructions can be generalised by 
introducing a gauge-invariant Goldstino superfield of the form 
\bea
{\mathfrak V}_n := {\mathfrak V} \frac{(\cD W)^{4n}}
{\big[ \cD^2 W^2 \bar \cD^2 \bar W^2 \big]^n}~,
\eea
for some integer $n$. The superfield \eqref{2.17} correspond to $n=1$.
It is obvious that ${\mathfrak V}_n $ obeys the constraints \eqref{2.6}.
Moreover,  ${\mathfrak V}_n $ is super-Weyl invariant. The superfields
 ${\mathfrak V} $ and  ${\mathfrak V}_n $ coincide if the gauge prepotential 
$V$  is chosen to be  ${\mathfrak V}$.
New FI-type invariants are obtained by making use of ${\mathfrak V}_n $
instead of ${\mathfrak V}$ in \eqref{2.25},
\bea
{\mathfrak I}^{(n)}_{\rm FI} =   \int \rd^4 x \rd^2 \q  \rd^2 \bar{\q} \, E\,
\U  {\mathfrak V}_n~.
\label{2.27}
\eea


\section{U(1) duality invariant models and BI-type terms}

Ref.  \cite{KMcC} presented a general family of $\sU(1)$ duality invariant models
for a massless vector supermultiplet coupled to off-shell supergravity, 
old minimal or new minimal.
Such a theory is described by a super-Weyl invariant action of the form
\bea
S_{\rm SDVM}[V;\U] =
\frac{1}{2}  \int \rd^4 x \rd^2 \q   \, \cE\,
W^2 &+&
\frac14  \int \rd^4 x \rd^2 \q  \rd^2 \bar{\q} \, E\,
\frac{W^2\,{\bar W}^2}{\U^2}\,
\L\!\left(\frac{\o}{\U^2},
\frac{\bar \o}{\U^2}\right)~.
\label{3.1}
\eea
Here $\cE$ is the chiral density,  
$\o:=  \frac{1}{8} \cD^2 W^2$, and $\L(\o,\bar \o)$ is areal analytic 
function satisfying the  differential equation \cite{KT1,KT2}
\bea
{\rm Im} \,\Big\{ \G
- \bar{\o}\, \G^2
\Big\} = 0~, \qquad \quad
\G  := \frac{\pa (\o \, \L) }{\pa \o}~.
\label{3.2}
\eea
These self-dual dynamical systems
are curved-superspace extensions 
of the globally supersymmetric systems 
introduced in \cite{KT1,KT2}.
The curved superspace extension \cite{CF}, $S_{\rm SBI} [V]$, 
of the supersymmetric Born-Infeld action \cite{CF,BG}
corresponds to the choice
\bea
&&\L(\o,{\bar\o}) =
\frac{g^2}
{ 1 + \hf\, A \, + \sqrt{1 + A +\frac{1}{4} \,B^2} }~,
\quad
A=g^2(\o  + \bar \o)~, \quad 
B=g^2(\o - \bar \o)~,~~
\eea
with $g$ a coupling constant.

In flat superspace (which, in particular, corresponds to  $\U=1$),
the fermionic sector of \eqref{3.1} was shown \cite{KMcC}
to possess quite remarkable properties. 
Specifically, only under the additional restriction
\be
\L_{u{\bar u}}(0,0) =  3 \L^{3}(0,0) ~, 
\label{cond-for-AV}
\ee
the component fermionic action proves to coincide, 
modulo a nonlinear field redefinition, 
with the Volkov-Akulov action \cite{VA}. This ubiquitous appearance 
of the Volkov-Akulov action in such models was explained in  \cite{K-FI}.
If the FI term is added to the flat-superspace counterpart of \eqref{3.1}, 
then the auxiliary scalar $D$  develops a non-vanishing expectation value, 
in general, for its algebraic equation of motion has a non-zero solution.
As a result, the supersymmetry becomes spontaneously broken, 
and thus the photino action should be related to the Goldstino action, 
due to the uniqueness of the latter. 

In supergravity, the situation is analogous to the rigid supersymmetric case. 
Let us add a standard FI term to the vector multiplet action \eqref{3.1}
coupled to new minimal supergravity,
\bea
S= \frac{3}{\k^2} \int \rd^4 x \rd^2 \q  \rd^2 \bar{\q} \, E\,
{L}\, {\rm ln} {L} +
S_{\rm SDVM}[V;L] -2f \int \rd^4 x \rd^2 \q  \rd^2 \bar{\q} \, E\,L V~.
\eea
Here the first term is the supergravity action. In general, this system describes 
spontaneously broken supergravity. It suffices to consider the case of vanishing gauge 
field, which corresponds to $V =\mathfrak V$. Using the nilpotency conditions 
\eqref{2.6} and  relation \eqref{2.99}, one may show that
\bea
S_{\rm SDVM}[{\mathfrak V};L] =
\frac{1}{2}  \int \rd^4 x \rd^2 \q   \, \cE\,
{\mathfrak W}^2 &+&
\frac14  \int \rd^4 x \rd^2 \q  \rd^2 \bar{\q} \, E\,
\frac{{\mathfrak V} (\cD {\mathfrak W})^3}{L^2}\,
\L\!\left(\frac{ \z}{L^2},
\frac{\z}{L^2}\right)~,
\label{3.6}
\eea
where $\z = -\frac18 (\cD {\mathfrak W})^2$. The auxiliary field may be eliminated by 
requiring the functional 
\bea
S_{\rm SDVM}[{\mathfrak V};L] -2f \int \rd^4 x \rd^2 \q  \rd^2 \bar{\q} \, E\,L 
{\mathfrak V}
\eea
 to be stationary under local rescalings 
$\d  {\mathfrak V} =\r {\mathfrak V} $, with $\r$ an arbitrary real superfield
(compare with \cite{KMcAT-M}).
The resulting algebraic equation proves to coincide with the one 
derived in \cite{K-FI}.

An important property of the standard FI term, which was pointed out in \cite{ADM},
is that it remains invariant under the second nonlinearly 
realised supersymmetry of the rigid supersymmetric Born-Infeld 
action \cite{BG}. This property implies the supersymmetric Born-Infeld 
action deformed by a FI term still describes partial $\cN=2 \to \cN=1$ supersymmetry
breaking \cite{K-FI,DFS}.
As for the novel FI-type terms \eqref{2.24} and \eqref{2.25}, 
they do not appear to share this fundamental property. 

${}$

\noindent
{\bf Acknowledgements:}  I am grateful to Ian McArthur and
Gabriele Tartaglino-Mazzucchelli for comments on the manuscript. 
The research presented in this  work is supported in part by the Australian 
Research Council, project No. DP160103633.

\appendix 

\section{Super-Weyl transformations}

It was first realised by Howe and Tucker  \cite{HT} that 
the Grimm-Wess-Zumino algebra of covariant derivatives \cite{GWZ}  is
invariant under super-Weyl transformations
of the form 
\begin{subequations} 
\label{superweyl}
\bea
\d_\s \cD_\a &=& ( {\bar \s} - \hf \s)  \cD_\a + \cD^\b \s \, M_{\a \b}  ~, \\
\d_\s \bar \cD_\ad & = & (  \s -  \hf {\bar \s})
\bar \cD_\ad +  ( \bar \cD^\bd  {\bar \s} )  {\bar M}_{\ad \bd} ~,\\
\d_\s \cD_{\a\ad} &=& \hf( \s +\bar \s) \cD_{\a\ad} 
+\frac{\ri}{2} \bar \cD_\ad \bar \s \,\cD_\a + \frac{\ri}{2}  \cD_\a  \s\, \bar \cD_\ad
+ \cD^\b{}_\ad \s\, M_{\a\b} + \cD_\a{}^\bd \bar \s\, \bar M_{\ad \bd}~,
~~~~~~
\eea
\end{subequations}
accompanied by 
the following transformations of the torsion superfields
\begin{subequations} 
\bea
\d_\s R &=& 2\s R +\frac{1}{4} (\bar \cD^2 -4R ) \bar \s ~, \\
\d_\s G_{\a\ad} &=& \hf (\s +\bar \s) G_{\a\ad} +\ri \cD_{\a\ad} ( \s- \bar \s) ~, 
\label{s-WeylG}\\
\d_\s W_{\a\b\g} &=&\frac{3}{2} \s W_{\a\b\g}~.
\label{s-WeylW}
\eea
\end{subequations} 
Here the super-Weyl parameter $\s$ is a covariantly chiral scalar superfield,  $\bar \cD_\ad \s =0$.

A tensor superfield $\cT $ (with its  indices suppressed)
is said to be super-Weyl primary of weight $(p,q)$
if its super-Weyl transformation law is 
\bea
\d_\s \cT =\big(p\, \s + q\, \bar \s \big) \cT~,
\label{A.3}
\eea
for some parameters $p$ and $q$.


\begin{footnotesize}

\end{footnotesize}

\end{document}